\documentclass[aps,pre,onecolumn,showpacs,a4paper]{revtex4}
\input epsf 
\usepackage{graphicx} 
\usepackage{amsmath}
\usepackage{amssymb}
\usepackage{amscd}

\begin{document} 

\title{On the stochastic pendulum with Ornstein-Uhlenbeck noise}
\author{Kirone Mallick}
 \affiliation{Service de Physique Th\'eorique, Centre d'\'Etudes de Saclay,
 91191 Gif-sur-Yvette Cedex, France}
 \email{mallick@spht.saclay.cea.fr}
\author{Philippe Marcq}
 \affiliation{Institut de Recherche sur les Ph\'enom\`enes Hors \'Equilibre,
 Universit\'e de Provence,
 49 rue Joliot-Curie, BP 146, 13384 Marseille Cedex 13, France}
 \email{marcq@irphe.univ-mrs.fr}
\date{March  16, 2004}

\begin{abstract}
 We study a frictionless pendulum subject to multiplicative random noise.
 Because of  destructive interference 
 between the angular  displacement of the system  and the noise term,
 the  energy  fluctuations   are reduced  when the noise 
 has a non-zero 
 correlation  time. 
 We derive the long  time behavior of the pendulum in the case
 of Ornstein-Uhlenbeck noise by   a  recursive adiabatic elimination procedure.
 An  analytical expression  for the 
 asymptotic probability distribution function 
 of the energy   is obtained  and the 
 results agree with numerical simulations. Lastly, we 
 compare  our  method  to other approximation schemes.
\end{abstract}
\maketitle 

 \section{Introduction}

  The  behavior of a 
 nonlinear dynamical system is strongly  modified when  
 randomness is taken into account: noise can shift
 bifurcation  thresholds,  create new phases  (noise-induced
 transitions), or  even generate  spatial patterns
 \cite{vankampen,lefever,anishchenko,vandenb1,toral}. The interplay
 of noise with nonlinearity   gives rise   to a variety of  phenomena  
 that  constantly motivate new research   
 of  theoretical   and practical  significance. Stochastic
 resonance \cite{marchesoni1} and  biomolecular Brownian  motors 
  \cite{jprost,reimann}   are  celebrated  examples
 of   nonlinear random systems of  current interest. In particular,
 stochastic ratchets  have generated a renewed interest in the study of
 simple mechanical systems  subject to random interactions,
 the common ancestor
 of  such  models    being  Langevin's description   of  Brownian
 motion. Many  unexpected  phenomena  appear when one generalizes
 Langevin's equations to include, {\it e.g.}, inertial terms, nonlinearities,
 external (multiplicative) noise or 
  noise with  finite  correlation time;
  each of these new features  opens a  field of investigations
  that calls for  specific techniques or
  approximation schemes \cite{gardiner}.
 Nonlinear oscillators with parametric noise  are often
  used as paradigms for the study of these various effects
 and related mathematical  methods \cite{landaMc}.
The   advantage of such  models  is  that they have
 an appealing  physical interpretation and  appear as building blocks
 in many different fields;  they can be simulated
 on a computer or constructed  as  real  electronic or mechanical
 systems \cite{strato,fauve}.
  Moreover, the mathematical apparatus needed to analyze them
 remains  relatively elementary (as  compared 
 to the  perturbative field-theoretical  methods  required for 
  spatio-temporal systems \cite{munoz}) and   can be expected to yield
  exact  and rigorous results.

  In the present work, we study the motion of a frictionless pendulum
  with parametric noise, which can be physically interpreted
  as  a randomly vibrating suspension axis. We  show that
  the long time behavior of 
   a stochastic   pendulum driven by a  colored  noise
 {\it  with finite correlation time } 
   is  drastically different from that  
 of a   pendulum subject to white noise. Whereas the average
 energy of the white-noise pendulum is a linear function of time,
 that of the  colored-noise pendulum grows only as the the square-root of time.
 Our analysis  is based on a generalization of the averaging  technique that
 we have used previouly  for nonlinear oscillators subject to white noise:
 in \cite{philkir1,philkir2,philkir3}
 an effective dynamics for the action variable  of the system is  derived
 after integrating  out the  fast angular variable,
 and   is  then   exactly solved.
 However, this  averaging technique as such 
ceases to  apply  to systems with colored noise  when 
 the time scale of the   fast variable becomes smaller  than the 
 correlation time of the noise. Correlations
 between the fast variable  and the noise  
  modify the long  time  scaling behavior of the system and therefore  must
 be taken into account. We shall develop here  a   method
 that systematically  retains  these  correlation   terms 
 before the fast variable is  averaged  out.
 This  will allow  us   to  derive analytical expressions 
 for the asymptotic probability distribution function   (P.D.F.)
  of the energy of the stochastic pendulum, and to  deduce 
  the long time behavior of  the system.  Our analytical  results are   verified 
 by  numerical simulations.  Finally, we shall compare
 our method and results with  some  known  approximation  schemes  used for
 multivariate systems with colored noise. We  shall  show in particular 
 that  small correlation time expansions  cannot explain the 
 anomalous  diffusion 
 exponent when   truncated at  any  finite order.
  The   partial 
  summation of   Fokker-Planck  type terms, used to derive 
 a `best  effective Fokker-Planck equation' for colored noise
  \cite{lindcol1,lindcol2}
 leads to results that  agree    with ours. We emphasize that
 the noise considered in this work has a finite correlation time
 and  its auto-correlation function does not have  long time tails.

 This article is organized as follows. In section II, we 
  analyse  the case where the parametric fluctuations  of the pendulum
  are  modeled by  Gaussian white noise, 
and   explain heuristically why colored noise leads to 
anomalous scaling of the energy.
 In section III, we study the case of Ornstein-Uhlenbeck noise
 and  explain how a recursive adiabatic elimination
 of the fast variable can be performed.  This allows us to derive
analytical results  for the P.D.F. of the energy, 
which we validate  with direct numerical simulations. 
In section IV, we compare  our results with effective Fokker-Planck
 approaches. Concluding remarks are presented in section V.

 \section{The pendulum with parametric noise}

    The dynamics of  a non-dissipative
 classical pendulum with parametric noise can be 
 described by the following system of stochastic differential equations
\begin{eqnarray}
     \dot \Omega  &=& -(\omega^2 + \xi(t))\,  \sin\theta   \, ,   
 \label{dynOmega} \\
      \dot \theta  &=&   \Omega   \,  , \label{dyntheta}  
 \end{eqnarray}
where $\theta$ represents the angular displacement and 
$\Omega$ the angular velocity.
 The   energy $E$  of the system is 
  \begin{equation}
 E  =  \frac{\Omega^2}{2}  - \omega^2 \cos \theta  \, .
 \label{defenergie}
 \end{equation}
 Equations~(\ref{dynOmega}-\ref{dyntheta})  describe the
 motion of a pendulum  of frequency  $\omega$ 
 whose suspension point is subject
 to a stochastic  force proportional to  the random function $\xi(t)$.
 Our aim is to study
 how the stochastic properties of  $\xi(t)$ are transfered to
 the dynamical variables $(\theta, \Omega)$ through a
 multiplicative and nonlinear coupling. We are chiefly interested
  in the case where  $\xi(t)$  has a non-zero correlation time, but
 we first   consider   Gaussian white noise. 
Eqs.~(\ref{dynOmega}-\ref{dyntheta}), as well as all the stochastic
differential equations below, are interpreted according to 
the rules of Stratonovich calculus.

\subsection{The white noise case}
\label{sec:white}

  The dynamics  of an oscillator subject to  multiplicative white noise
 has been studied by a number of investigators
  \cite{strato,  bourret, vankamp2,  lindmoment}.
 In particular, Lindenberg  {\it et al.} have shown \cite{lindparam}
 that the physical origin of  the energetic instability of  a stochastic oscillator
 can be quantitatively related to  
 the parametric resonance  of  the  underlying deterministic oscillator. 
 In this section,  an  instability of the same type 
  is found for the stochastic pendulum 
 subject to parametric white noise,  using 
 the  adiabatic averaging method.

\begin{figure}[th]
 \hfill
 \includegraphics*[width=0.4\textwidth]{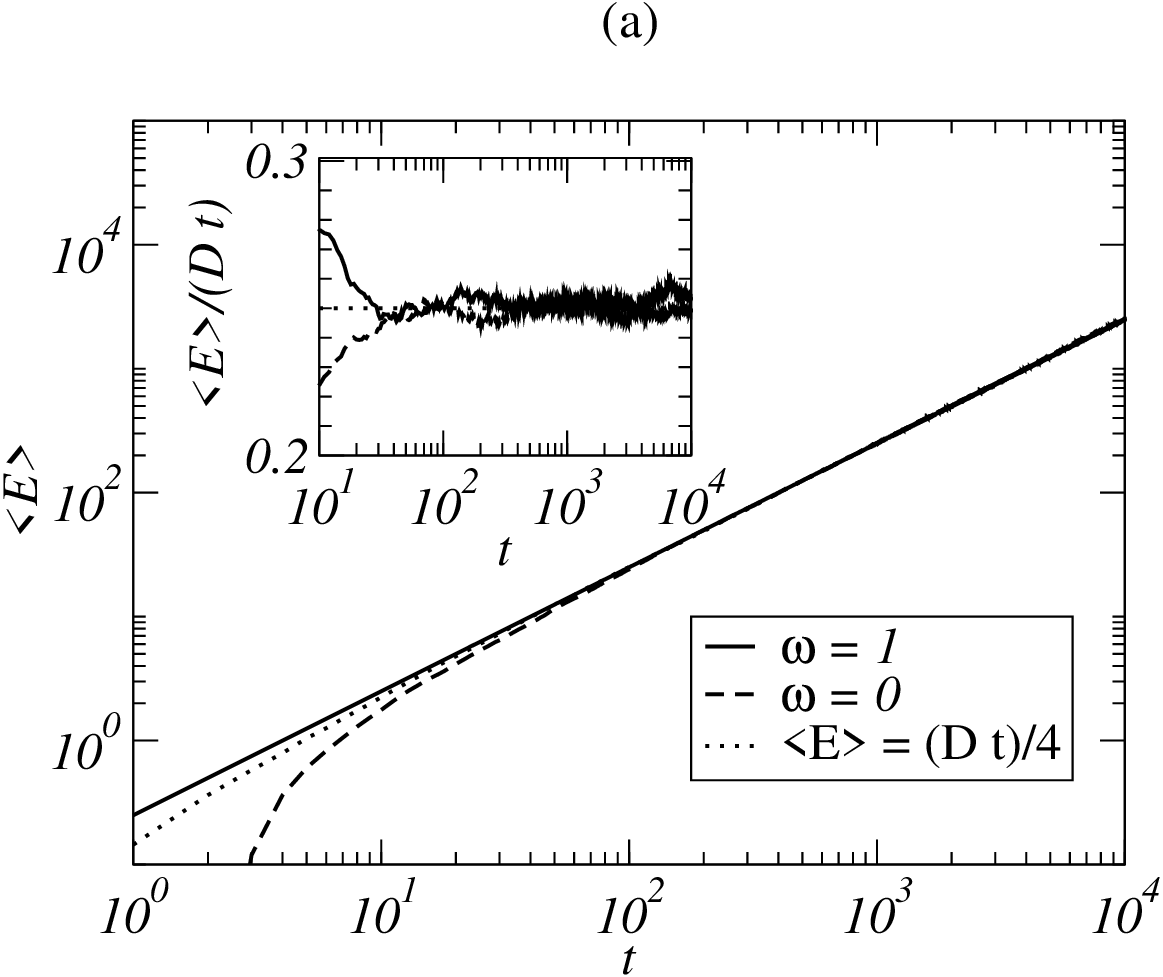}
 \hfill
 \includegraphics*[width=0.38\textwidth]{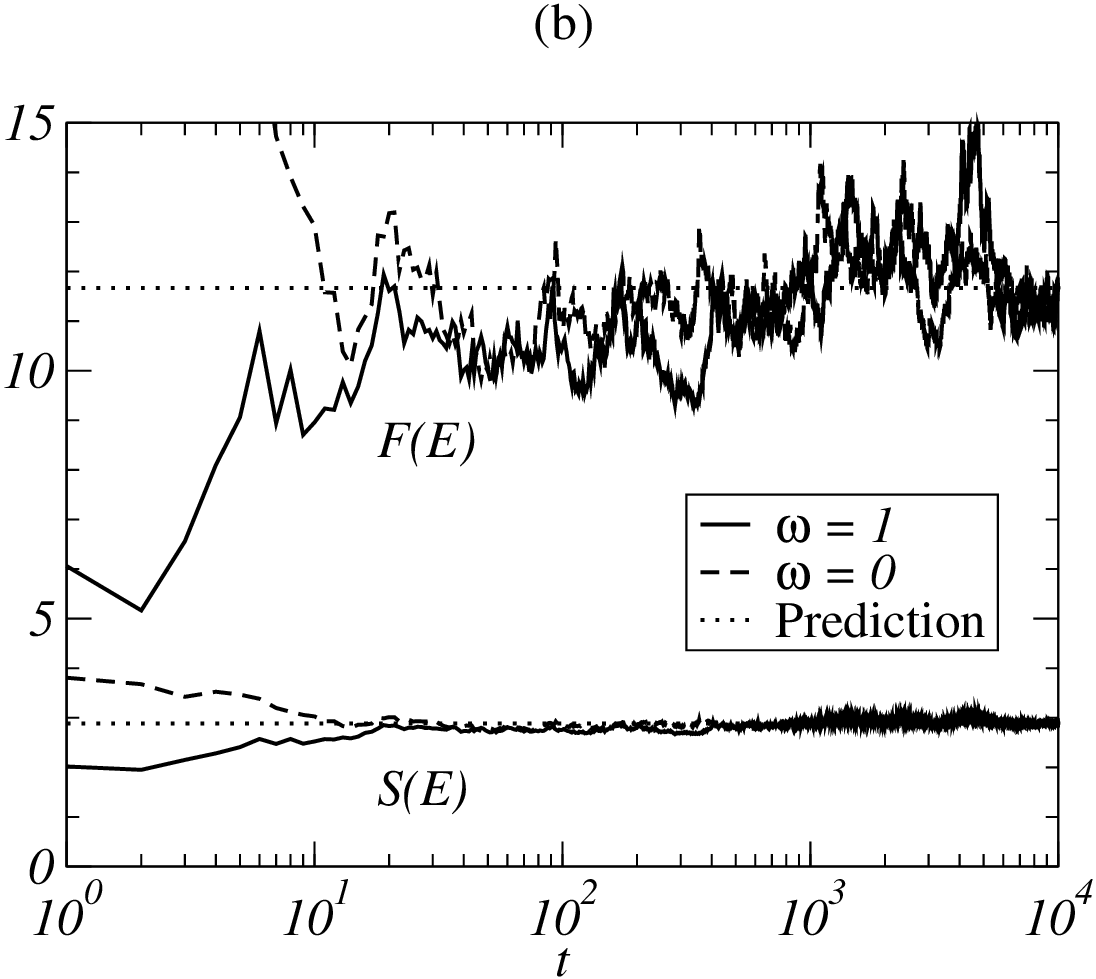}
 \hfill

    \caption{\label{fig:blanc} Stochastic pendulum with Gaussian white
noise: Eqs.~(\ref{dynOmega}-\ref{dyntheta}-\ref{defbruitblanc}) 
are integrated numerically for ${\mathcal D} = 1$. Ensemble averages 
are computed over $10^4$ realizations. For numerical values of the pulsation
$\omega = 1.0$ and $0.0$, we plot: $(a)$ the average 
$\langle E \rangle$ and the ratio $\langle E \rangle/(\mathcal{D} \,t)$ 
(inset),  $(b)$ the skewness and flatness factors of $E$ \emph{vs.}  time $t$. 
The asymptotic behavior of  these  observables   agrees with 
Eqs.~(\ref{eq:scalEblanc}--\ref{eq:flatEblanc}) 
(dotted lines in the figures), irrespective of the value of $\omega$.
}
\end{figure}

Let  $\xi(t)$  be  a Gaussian white noise
 of zero mean value and amplitude ${\mathcal D}$:
\begin{eqnarray}
       \langle \xi(t)  \rangle &=&   0   \, ,\nonumber \\
   \langle \xi(t) \xi(t') \rangle  &=&  {\mathcal D} \, \delta( t - t') .
   \label{defbruitblanc}
 \end{eqnarray}
 Using  elementary dimensional analysis, we notice  from Eq.~(\ref{dynOmega})
 that  $ \dot \Omega \sim \xi$, {\it i.e.},   the angular velocity 
 $\Omega$ grows  as $t^{1/2}$  and therefore 
 $\theta \sim t^{3/2}$.
 This observation
 can be put on a stronger basis by using  the Fokker-Planck 
 equation, associated with Eqs.~(\ref{dynOmega}) and (\ref{dyntheta}),  that
 describes the evolution of the Probability Distribution Function
 $P_t(\theta, \Omega)$:
\begin{equation}
\frac{ \partial   P_t }{\partial t}  = 
 -\frac{ \partial   }{\partial \theta} \left( \Omega  P_t \right) 
 +   \frac{ \partial   }{\partial \Omega}
  \left( \omega^2 \sin \theta  P_t \right) 
 + \frac{ {\mathcal D}}{2} \sin^2\theta \frac{ \partial^2   P_t }
 {\partial \Omega^2}   \,.
\label{FPblanc}
\end{equation}
 From  Eq.~(\ref{dyntheta}), we observe that the angular
 variable $\theta$ varies rapidly as compared to $\Omega$.
 Thus, following \cite{philkir1}, we     
  assume  that,  in the long time limit, the angle $\theta$
  is uniformly  distributed   over  $[0, 2\pi]$.
 This allows  us to average   Eq.~(\ref{FPblanc})  
  over   the angular variable  and to derive an effective Fokker-Planck 
 equation  for the marginal distribution  $\tilde{P}_t(\Omega)$:
  \begin{equation}
\frac{ \partial  {\tilde P_t} }{\partial t}  = 
\frac{ {\mathcal D} }{4}  \;
 \frac{ \partial^2  {\tilde P_t}   } {\partial \Omega^2}      \, ,
\label{moyblanc} 
  \end{equation}
 where we have replaced   ${\sin^2\theta}$  by its mean value  1/2.
 From this effective  Fokker-Planck  equation, we readily deduce that
$\Omega$ is a Gaussian variable with P.D.F. 
\begin{equation}
  \label{eq:pdfOmegablanc}
  {\tilde P_t}(\Omega) = \frac{1}{\sqrt{\pi \mathcal{D} t}} \;
\exp{\left( -\frac{\Omega^2}{\mathcal{D} t} \right)}.
\end{equation}
We thus recover that $\Omega$ grows as $t^{1/2}$ when  $t \to \infty$.
 Because  $E \simeq \Omega^2/2$ (up to a term
 that remains  bounded), we deduce the 
P.D.F. of the energy 
\begin{equation}
  \label{eq:pdfEblanc}
  {\tilde P_t}(E) =  \sqrt{\frac{2 }{\pi  \mathcal{D} t}} \;
   E^{ -\frac{1}{2}} \exp{\left( -\frac{2 E}{\mathcal{D} t} \right)}.
\end{equation}
 From Eq.~(\ref{eq:pdfEblanc}), we obtain the scaling 
behavior of the average energy 
\begin{equation}
  \label{eq:scalEblanc}
  \langle E \rangle = \frac{ {\mathcal D} }{4}\;  t \, ,
\end{equation}
and also  the skewness and flatness factors 
\begin{eqnarray}
  S(E) &=& \frac{\langle E^3 \rangle}{\langle E^2 \rangle^{3/2}}
= \frac{\Gamma(\frac{7}{2}) \; \Gamma(\frac{1}{2})^{1/2}}{\Gamma(\frac{5}{2})^{3/2}}
=  \frac{5}{\sqrt{3}}  \simeq 2.887\ldots,   \label{eq:skewEblanc}\\
  F(E) &=& \frac{\langle E^4 \rangle}{\langle E^2 \rangle^2}
= \frac{\Gamma(\frac{9}{2}) \; \Gamma(\frac{1}{2})}{\Gamma(\frac{5}{2})^2}
= \frac{35}{3} \, ,   \label{eq:flatEblanc}
\end{eqnarray}
where $\Gamma()$  is the Euler Gamma function.
  We conclude from Eq.~(\ref{eq:scalEblanc}) that   
   the average energy of a
 frictionless  pendulum with white parametric noise 
 grows linearly with time. 
These results, Eqs.~(\ref{eq:scalEblanc}-\ref{eq:flatEblanc}), 
   agree  with   numerical simulations (see Fig.~\ref{fig:blanc}).

\subsection{Scaling analysis for colored noise}

 When the noise $\xi(t)$ is colored, it is not possible to write
 a closed equation for the P.D.F. $P_t(\theta, \Omega)$.  However, 
   the scaling behavior of the dynamical variables can  be deduced from a 
  self-consistent reasoning similar to that used in \cite{philkir2, philkir3}.
 Suppose {\it a priori} that  we have the following scaling behavior 
 in the long time limit 
 \begin{equation}
\Omega \sim t^{\alpha}   \, , 
 \label{defalpha} 
 \end{equation}
 where $\alpha$
 is an unknown exponent to be determined.  We find  from Eq.~(\ref{dyntheta})
 that  $\theta \sim  t^{\alpha + 1}$,
 and Eq.~(\ref{dynOmega}) 
 can then be   written as 
 \begin{equation}
  \dot\Omega \simeq  \xi(t) \,   \sin t^{\alpha + 1}\,   .
 \end{equation} 
  (We could have retained  the deterministic term $-\omega^2 \sin\theta$ but  it 
 would  not affect this scaling analysis.) We now  take
   $\xi(t)$  to  be  a discrete  dichotomous
 noise with  correlation time $\tau$   and with  values 
 $ \pm 1$. The previous  equation, then, becomes
 \begin{equation}
  \Omega(t) \sim \sum_{k =1}^{t/\tau} \epsilon_k \int_{(k-1)\tau}^{k\tau} 
  \sin x^{\alpha + 1} {\rm d}x  \, , \,\,\,\,\, \hbox{ with  }
  \epsilon_k = \pm 1   \, .
 \end{equation}
 We estimate the last integral  by  an  integration  by parts:
 \begin{equation}
(\alpha + 1) \int_{t}^{t +\tau}x^{\alpha}
 \frac{ \sin x^{\alpha + 1} }{  x^{\alpha}} {\rm d}x  = 
\left[ - \frac{\cos(x^{\alpha + 1})}{x^{\alpha}}
 \right]_{t}^{t+\tau}
    + {\mathcal O}(t^{- \alpha - 1}) \, ,
\end{equation} 
  and   obtain  
\begin{equation}
 \langle \Omega^2 \rangle \sim \sum_{k =1}^{t/\tau} 
 \left( \int_{(k-1)\tau}^{k\tau}  \sin x^{\alpha + 1} {\rm d}x  \right)^2
   \sim \sum_{k =1}^{t/\tau} \frac{1}{ (k\tau)^{2\alpha}} 
 \sim t^{ 1 - 2 \alpha} \,.
\end{equation}
This result is compatible with the {\it a priori} scaling Ansatz
 (\ref{defalpha})  only for 
 $\alpha = 1/4$.  We thus conclude  from this qualitative argument 
   that,  in  the presence of colored
 noise, the   energy $E$  of the system defined in Eq.~(\ref{defenergie})  
  grows as the square-root of time 
(as opposed to the linear growth obtained for white noise).
  A non-zero  correlation time of the noise, however small, 
   modifies the long time  scaling behavior of the system.

  In the next section, we  develop a method to 
 put  this  qualitative analysis  on a firm basis and     derive   precise
 analytic expressions that can be compared 
 quantitatively with  numerical results.

\section{The averaging method for Ornstein-Uhlenbeck noise}
\label{sec:color}

   From now on, we consider   the random noise $\xi$ 
   to be  an Ornstein-Uhlenbeck
 process, {\it i.e.},  a Gaussian  colored noise with correlation function
 given by:
\begin{equation}  
 \langle \xi(t) \xi(t') \rangle  =  
\frac{\mathcal D}{2 \, \tau}   \, {\rm e}^{-|t - t'|/\tau} \,,
\label{correlationOU}
\end{equation}
where $\tau$ is the correlation time of the noise. This noise $\xi$ can
 be generated  from  white noise via  the  Ornstein-Uhlenbeck equation
 \begin{equation}  
      \dot \xi   =  -\frac{1}{\tau} \xi  +  \frac{1}{\tau} \eta(t) \, ,
\label{eqOU}
\end{equation}
 where $\eta(t)$ is a white noise of auto-correlation
 function  ${\mathcal D} \, \delta( t - t')$. In the stationary
 limit, $ t, t' \gg \tau , $ the solution of Eq.~(\ref{eqOU}) 
 satisfies Eq.~(\ref{correlationOU}). The 
 pendulum with Ornstein-Uhlenbeck noise is  thus  written
as a three-dimensional stochastic dynamical system coupled to  a white noise
 $\eta(t)$:
\begin{eqnarray}
     \dot \Omega  &=& - \omega^2 \, \sin\theta  - \xi\,  \sin\theta   \, ,   
 \label{dyncolOmega} \\
      \dot \theta  &=&   \Omega   \,  , \label{dyncoltheta}    \\
  \dot \xi   &=&  -\frac{1}{\tau} \xi  +  \frac{1}{\tau} \eta(t) \label{OU} \, . 
 \end{eqnarray}
The Fokker-Planck equation   for the  three-dimensional 
  P.D.F. $P_t(\theta, \Omega, \xi)$
is given by
\begin{equation}
\frac{ \partial   P_t }{\partial t}  = 
 -\frac{ \partial   }{\partial \theta} \left( \Omega  P_t \right) 
 +   \frac{ \partial   }{\partial \Omega}
  \left( (\omega^2  + \xi)  \sin \theta  P_t \right) 
+  \frac{1}{\tau}  \frac{ \partial   }{\partial \xi } \left(  \xi  P_t \right) 
 + \frac{ {\mathcal D}}{2 \tau^2 } \frac{ \partial^2   P_t }
 {\partial \xi^2}   \,.
\label{FPcolor}
\end{equation}
 For  colored noise, there is no  closed Fokker-Planck equation  
 for the original  P.D.F. on phase space, $P_t(\theta, \Omega)$,   and  
 only approximate  evolution equations can be written. 
 We shall not use any  effective dynamics  to derive our
 results  but rather start with the exact three-dimensional 
 Fokker-Planck equation~(\ref{FPcolor}) from which we shall   integrate 
 out the fast variable.

   \subsection{Zeroth-order averaging}

 We    show  here that the  averaging procedure used in  
 section  \ref{sec:white}  for white noise   leads to erroneous results
 for colored noise. 
  Averaging  the  Fokker-Planck  equation~(\ref{FPcolor})
 over the fast angular variable
 $\theta$, we find that the marginal  P.D.F. 
 $ {\tilde P_t}(\Omega, \xi)$ obeys the Ornstein-Uhlenbeck diffusion
 equation 
 \begin{equation}
\frac{ \partial {\tilde P_t} }{\partial t}  = 
 \frac{1}{\tau}  \frac{ \partial   }{\partial \xi } \left(  \xi {\tilde P_t} \right) 
 + \frac{ {\mathcal D}}{2 \tau^2 } \frac{ \partial^2  {\tilde P_t} }
 {\partial \xi^2}   \, ;
\end{equation}
  the variable $\Omega$  no more 
  appears    in this  averaged Fokker-Planck  equation
 and the associated   stochastic two-dimensional system reads 
 \begin{equation}
\dot \Omega  = 0    \,\, \,\, \,\, \,\, \hbox{ and }   \,\, \,\,
   \dot \xi  =  -\frac{1}{\tau} \xi  +  \frac{1}{\tau} \eta(t) \, .
\label{average0}
\end{equation}
 In this averaged system, 
 the angular velocity  $\Omega$ is no more  stochastic 
 and,   even worse, it   is constant in time. 
 By  integrating  out the fast variable  
 $\theta$  without taking into account the correlations
  between  $\theta$  and $\xi$, 
 the dynamical variable $\Omega$  has been decoupled
  from the noise;   in other words,
  the noise itself has been averaged out of the   system.
 In fact,  when the  typical  variation time   of 
 $\theta$ ({\it i.e.} the time during which $\theta$ varies by 2$\pi$) 
 becomes less than $\tau$,  the noise $\xi$ is  roughly constant
 during a period of $\sin\theta$. 
 Thus, if  $\theta$ and $\xi$ are (wrongly) treated as independent variables,
  $(\sin \theta)\, \xi$ is  averaged to $0$  at the leading order.
  This problem did not occur in section~\ref{sec:white} where 
   $\xi$ was   a  white noise. In that case,
 the rapid fluctuations of the phase do not  wipe out
 the noise, and  $(\sin \theta)\, \xi$ is  averaged to  
$\xi/\sqrt{2}$ to yield Eq.~(\ref{moyblanc}). 

  In the next subsections, we develop an   averaging scheme
 that allows us  to  eliminate adiabatically the fast variable while retaining
   the  correlation terms.  
 The idea is to define  recursively  a new set of dynamical variables
 that embodies   the  correlations order by order. This will  enable us 
 to derive sound asymptotic results for the pendulum with
 colored noise.  In this scheme, 
Eq.~(\ref{average0}) appears  as a zeroth order
 approximation, and its correct interpretation is not that 
  $\Omega$ is conserved but that its variations are slower than
  that of  normal diffusion.

 \subsection{First-order averaging}
\label{sec:1staveraging}

  Multiplying     both sides of Eq.~(\ref{dyncolOmega})  by $\Omega$,  
  and using  Eq.~(\ref{dyncoltheta}), we  obtain   
\begin{equation}
\Omega  \dot \Omega  =  
 - \Omega \left( \omega^2 \, \sin\theta  + \xi\,  \sin\theta \right)  =  
- \omega^2 \dot\theta \sin\theta - \xi\, ( \dot\theta \sin\theta ) \,.
\end{equation}
  Introducing the  energy $E$ of the system defined in
 Eq.~(\ref{defenergie}), this equation becomes 
\begin{equation}
 \frac{\rm{d} E }{\rm{d}t } =   \frac{\rm{d}  }{\rm{d}t } 
\left( \frac{\Omega^2}{2}  - \omega^2 \cos \theta\right)
  =  - \xi\, ( \dot\theta \sin\theta ) \, .
\end{equation}
 We  now  transform the  right hand side   by writing it  as a total
 derivative plus a correction term: 
\begin{equation}
  \frac{\rm{d} E }{\rm{d}t } = \frac{\rm{d}  }{\rm{d}t } 
 \left(  \xi\, \cos \theta\right)  - \dot   \xi  \, \cos \theta \, .
\end{equation} 
 Using  Eq.~(\ref{OU}),   we obtain 
\begin{equation}
 \frac{\rm{d}  }{\rm{d}t } \left( E    -   \xi\, \cos \theta \right)
 = \frac{ \cos \theta  } {\tau} \,\xi -  \frac{ \cos \theta  } {\tau} \, \eta 
    \, . 
\end{equation}  
 This leads us to define  a new dynamical variable  $E_1$
\begin{equation}
  E_1 =  E    -   \xi\, \cos \theta = 
  \frac{\Omega^2}{2}  - (\omega^2 + \xi) \cos \theta   \, ,
 \label{defE1}
\end{equation} 
and  to rewrite the stochastic system 
  (\ref{dyncolOmega}, \ref{dyncoltheta}, \ref{OU})
  in terms of the set of variables $(E_1, \theta, \xi)$:
\begin{eqnarray}
     \dot E_1  &=&    \frac{ \cos \theta  } {\tau} \,\xi 
  -  \frac{ \cos \theta  } {\tau} \, \eta(t)   \, ,   
 \label{dyncolE1} \\
      \dot \theta  &=&   \Omega\left(  E_1, \theta, \xi   \right)
= \sqrt{2  \Big(E_1  +  (\omega^2 + \xi)\cos \theta\Big)  }
    \,  , \label{dyncol2theta}    \\
  \dot \xi   &=&  -\frac{1}{\tau} \xi  +  \frac{1}{\tau} \eta(t) 
 \label{OU2}\, .
 \end{eqnarray}

\begin{figure}[th]
\centerline{\includegraphics*[width=0.4\textwidth]{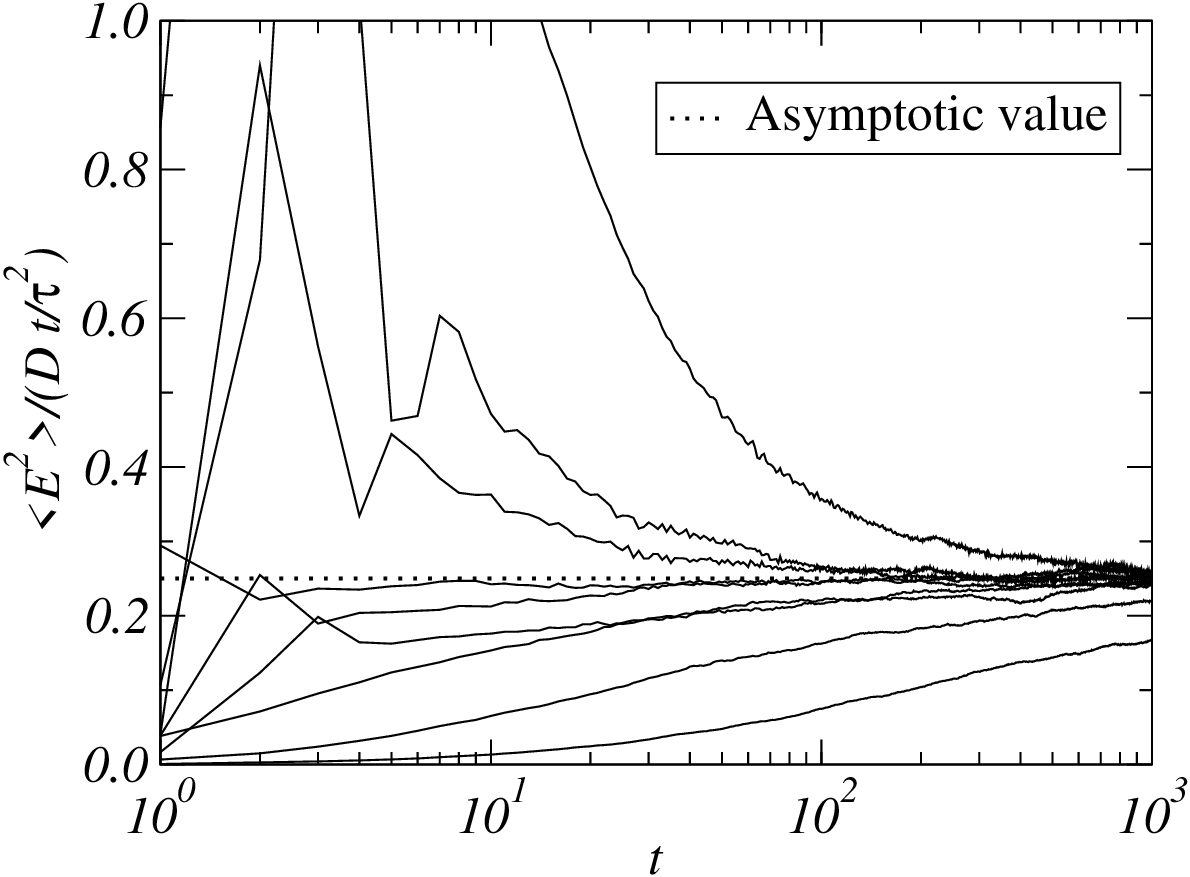}}

    \caption{\label{fig:collapse} Stochastic pendulum with 
Ornstein-Uhlenbeck colored noise: 
Eqs.~(\ref{dyncolOmega}--\ref{OU}) 
are integrated numerically for $\omega = 1.0$ and
$({\mathcal D}, \tau) = (1, 0.1), (1, 1), (1, 10),
(10, 0.1), (10, 1), (10, 10), (100, 0.1), (100, 1), (100, 10)$. 
Ensemble averages are computed over $10^4$ realizations. We plot the ratio 
$\langle E^2 \rangle/(\mathcal{D} \,t/\tau^2)$ \emph{vs.}  time $t$. 
The dotted line in the figure corresponds to   the asymptotic 
value $0.25$ predicted by second-order averaging 
(Eq.~(\ref{eq:scalE2color})).  Note that convergence is slower
for smaller values of the correlation time $\tau$.
}
\end{figure}

The advantage of this system as compared to the previous one
 (\ref{dyncolOmega}, \ref{dyncoltheta}, \ref{OU})   is that 
 the white noise $\eta(t)$  now  appears in the equation for the
 dynamical variable  $E_1$ and this white noise contribution will survive
 the averaging process. The Fokker-Planck equation 
  for the  P.D.F. $P_t( E_1,\theta,    \xi)$  associated with
 Eqs.~(\ref{dyncolE1}, \ref{dyncol2theta} and \ref{OU2}) reads:
\begin{equation}
\frac{ \partial   P_t }{\partial t}  = 
 -   \frac{ \partial   }{\partial E_1}
  \left(\frac{ \cos \theta  } {\tau} \,\xi   P_t \right) 
 -\frac{ \partial   }{\partial \theta} \left( \Omega ( E_1, \theta, \xi )
 P_t \right) 
+  \frac{1}{\tau}  \frac{ \partial   }{\partial \xi } \left(  \xi  P_t \right) 
 + \frac{ {\mathcal D}}{2 \tau^2 }  \left\{
   \cos^2 \theta  \frac{ \partial^2   P_t   }{\partial E_1^2}
   -  2 \cos \theta \frac{ \partial^2   P_t   }{\partial E_1\partial \xi   }
+ \frac{ \partial^2   P_t } {\partial \xi^2}  \right\}           \,.
\label{FPcolor2}
\end{equation}
Averaging this equation with respect to  $\theta$ leads to
 the following evolution equation for the marginal
 distribution $ {\tilde P_t}(E_1, \xi)$:
 \begin{equation}
\frac{ \partial {\tilde P_t} }{\partial t}  =  
 \frac{1}{\tau}  \frac{ \partial   }{\partial \xi } \left(  \xi {\tilde P_t} \right)
 +  \frac{ {\mathcal D}}{4 \tau^2 } \frac{ \partial^2  {\tilde P_t} }
 {\partial E_1^2}
 + \frac{ {\mathcal D}}{2 \tau^2 } \frac{ \partial^2  {\tilde P_t} }
 {\partial \xi^2}   \,.
\label{FPav1}
\end{equation}
 The variable  $E_1$  now appears   in the averaged 
 Fokker-Planck equation, and the associated stochastic two-dimensional
 system reads 
\begin{eqnarray}
\dot  E_1    &=&     \frac{1}{\sqrt{2}\tau} \eta_1(t) \,, \label{average1} \\
\dot \xi   &=&  -\frac{1}{\tau} \xi  +  \frac{1}{\tau} \eta(t) \,, 
 \label{OUaverage1}
\end{eqnarray}
where $\eta(t)$ and  $\eta_1(t)$ are two independent white noises 
 of amplitude ${\mathcal D}$.  This 
 effective  dynamics  is exactly solvable (the variables 
 $E_1$ and $\xi$  are  decoupled thanks to  the absence 
 of the second-order cross-derivative term in Eq.~(\ref{FPav1})).
 If we compare this system with
 the one obtained by naive averaging (\ref{average0}), we observe that
 the dynamical variable $E_1$ (and therefore the energy
 of the system)  is no longer constant, 
 rather it   grows as the
 square-root of time. When $ t \to \infty$,  $E_1$  becomes  identical to
 the energy of the system (up to terms that  remain finite).
 We therefore determine the long  time statistics of the energy from 
  Eq.~(\ref{average1}) by identifying $E_1$ to $E$  and by 
 imposing the  physical condition $E \ge 0$. The energy is thus a Wiener
 process on a half-line and its P.D.F. is given by
 \begin{equation}
   P_t(E) = \frac{2 \tau  } {\sqrt{ \pi{\mathcal D} t }} 
      \exp\left( - \frac{ \tau^2 E^2}{ {\mathcal D} t }   \right)
 \,\,\,\, \hbox{ with } \,\,\,\, E  \ge 0  \,.  
  \label{PDFav1} 
\end{equation} 
 From this expression we calculate  the first two moments of the energy 
  \begin{eqnarray} 
  \langle E  \rangle &=&\frac{1}{ \sqrt{\pi} } 
  \left( \frac{ {\mathcal D}  t }{ \tau^2 }\right)^ \frac{1}{2}  
 \simeq 0.564
  \left( \frac{ {\mathcal D}  t }{ \tau^2 }\right)^ \frac{1}{2}  \,, 
  \label{diff1}  \\
   \langle E^2  \rangle &=& \frac{1}{2}
  \frac{ {\mathcal D}  t }{ \tau^2 } \, .
  \label{quad1}
  \end{eqnarray} 

 These  expressions provide   scaling  relations
  between the averages, the time $t$
  and   the dimensional  parameters  of the problem,
 $ {\mathcal D}$ and $\tau$.
  We have  verified numerically that these
  scalings are correct (see Fig.~\ref{fig:collapse}). 
  However, the prefactors
  that appear  in Eqs.~(\ref{diff1}) and (\ref{quad1}) 
   are  pure numbers and     do  not
   agree with  the results of our  numerical simulations.
   We conclude that Eq.~(\ref{diff1}) is  exact at leading order 
  as it  gives 
 the correct asymptotic  scaling for the   
 energy, $E \propto t^{1/2}$   
 but fails to  provide the prefactors.
 The reason is that some correlations
 between $\theta$ and $\xi$ have still  been neglected in the averaging
 procedure. Carrying out the calculations
  to the next higher  order will enable  us
 to derive the correct expressions for  the prefactors.

\subsection{Second-order averaging}

  More precise results can  indeed be  derived by applying recursively
 the  procedure  described above. In the Langevin equation~(\ref{dyncolE1}) for
  $E_1$, we perform one more  `integration by parts' and obtain 
   \begin{equation}
     \dot E_1  =     \frac{ \xi  \cos \theta  } {\tau } \,
\frac{  \dot\theta  } {\Omega } 
  -  \frac{ \cos \theta  } {\tau} \, \eta(t)   =  \frac{\rm{d}  }{\rm{d}t }
 \left(   \frac{  \xi  \sin  \theta  } {\tau\Omega }  \right)
   - \frac{    \sin  \theta  } {\tau\Omega } \dot\xi  + \frac{  \dot \Omega }
  {\Omega^2 }  \frac{    \sin  \theta  } {\tau } \,\xi
 -  \frac{ \cos \theta  } {\tau} \, \eta(t) \, .
 \label{dyncolU2} 
 \end{equation}
   This leads us to  introduce  a   new variable $E_2$ defined as
 \begin{equation}
  E_2   =  E_1 -  \frac{ \xi   \sin  \theta  } {\tau\Omega } 
   =  \frac{\Omega^2}{2}  - (\omega^2 + \xi) \cos \theta 
  -  \frac{ \xi \sin  \theta  } {\tau\Omega }        \, .
\label{defZ} 
 \end{equation}
 Using Eqs.~(\ref{dyncolOmega}) and (\ref{OU}), 
  Eq.~(\ref{dyncolU2}) becomes 
  \begin{equation}
   \dot  E_2   =  -  \frac{ \xi  \sin  \theta  } {\tau^2 \Omega } 
 - \frac{ \xi  \sin^2  \theta  }  { \tau \Omega^2 } \,(\omega^2 + \xi )
 -  \left( \cos \theta  +  
  \frac{   \sin  \theta  } {\tau \Omega  }  \right) 
  \frac{ \eta(t) } {\tau} \, . 
\label{dynZ0}
\end{equation}
 In this equation we must express  the variable $\Omega$ in terms
 of $E_2$, $\theta$ and $\xi$. Inverting 
  the  relation~(\ref{defZ}),  we deduce that 
  \begin{equation}
  \Omega = \left(  2E_2 \right)^{ \frac{1}{2}  }  
  + (\omega^2 + \xi) \frac{    \cos \theta  } { (2E_2)^{ \frac{1}{2}  } }
 +  \frac{   \xi   \sin  \theta  } { \tau  (2E_2)} 
 +  {\mathcal O} \left( \frac{1}{ E_2^{ \frac{3}{2}} }   \right) \, ,
\label{OmgZ}
\end{equation}
 where we have retained  terms up to  the order $1/E_2$.
 From  Eq.~(\ref{dynZ0}),  we deduce  the Langevin equation for $E_2$  
  \begin{equation}
  \dot  E_2   =   J_E( E_2, \theta, \xi ) 
 + D_E( E_2, \theta, \xi )  \; \frac{ \eta(t) } {\tau} \,
  +  {\mathcal O} \left( \frac{1}{ E_2^{ \frac{3}{2}} }   \right) \, ,
\label{dyncolZ} 
 \end{equation}
where we have defined 
\begin{eqnarray}
 J_E( E_2, \theta, \xi )    &=&  
 - \frac{\xi \sin  \theta  } {\tau^2 (2E_2)^{ \frac{1}{2}} } 
 - \frac{  \xi \sin^2  \theta  }  { \tau (2E_2) } (\omega^2 + \xi)  \,, \\
  D_E( E_2, \theta, \xi )   &=&   -      \left(  \cos \theta  +  
  \frac{   \sin  \theta  } {\tau (2E_2)^{ \frac{1}{2}} }  \right)  \, . 
\end{eqnarray}
 This equation, combined with Eqs.~(\ref{dyncoltheta} and \ref{OU}),
 defines a three-dimensional stochastic system  for  the variables
  $(E_2, \theta, \xi)$.  The Fokker-Planck equation 
  for the  P.D.F. $P_t( E_2,\theta,    \xi)$ is 
\begin{eqnarray}
\frac{ \partial   P_t }{\partial t}  = 
 &-&   \frac{ \partial   }{\partial E_2}  \left( J_E  P_t \right) 
- \frac{ \partial }{\partial \theta} 
 \left( \Omega ( E_2, \theta, \xi ) P_t \right) 
+  \frac{1}{\tau}  \frac{ \partial   }{\partial \xi } \left(  \xi  P_t \right) 
 \nonumber \\
 &+& \frac{ {\mathcal D}}{2 \tau^2 }  \left\{
 \frac{ \partial   }{\partial E_2} D_E
 \frac{ \partial   }{\partial E_2} ( D_E P_t) 
+ \frac{ \partial^2   }{\partial \xi \partial E_2} ( D_E P_t) 
+\frac{ \partial   }{\partial E_2} D_E
 \frac{ \partial  P_t }{\partial \xi}
+ \frac{ \partial^2   P_t } {\partial \xi^2}  \right\}           \,.
\label{FPcolor3}
\end{eqnarray}
 We now integrate out the fast angular variable  $\theta$
 from Eq.~(\ref{FPcolor3}), retaining  only the leading  term in the
 average of the expression 
 $\frac{ \partial   }{\partial E_2} D_E
 \frac{ \partial   }{\partial E_2} ( D_E  P_t) ,$
 (recalling that  
  ${ \partial   }/{\partial E_2}$  scales as  $E_2^{-1}$,
  the contribution of the subdominant terms is 
  of the order of $E_2^{- 5/2}$  and is negligible  in the long time limit).
 We thus obtain the following evolution equation for the marginal
 distribution $ {\tilde P_t}(E_2, \xi)$:
 \begin{equation}
\frac{ \partial {\tilde P_t} }{\partial t}  =  
\frac{ \partial   }{\partial E_2}
 \left( \frac{ \omega^2  \xi +  \xi^2 } { 4\tau E_2} {\tilde P_t}   \right) 
+ \frac{1}{\tau}  \frac{ \partial   }{\partial \xi } \left(  \xi {\tilde P_t} \right)
 +  \frac{ {\mathcal D}}{2 \tau^2 } \left\{ \frac{1}{2}
 \frac{ \partial^2  {\tilde P_t} }
 {\partial E_2^2}
 +  \frac{ \partial^2  {\tilde P_t} } {\partial \xi^2}  \right\}    \,.
\label{FPav2}
\end{equation}
Although the  cross-derivative terms between $E_2$ and $\xi$ vanish,
 the  two variables are coupled through the drift term.
 From Eq.~(\ref{FPav2}) we  derive 
  an effective, two-dimensional,  stochastic  system in $E_2$ and $\xi$:
\begin{eqnarray}
\dot  E_2    &=& - \frac{ \omega^2  \xi +   \xi^2 } { 4\tau E_2} +
   \frac{1}{\sqrt{2}\tau} \eta_2(t) \,,
 \label{average2}    \\
\dot \xi   &=&  -\frac{1}{\tau} \xi  +  \frac{1}{\tau} \eta(t) \,, 
 \label{ZOU}
\end{eqnarray}
 $\eta(t)$ and  $\eta_2(t)$  being  two independent white noises 
 of amplitude ${\mathcal D}$.  If we compare this system deduced
 by a  second-order averaging with the one obtained at first order
 Eqs.~(\ref{average1},\ref{OUaverage1}), we observe that
  the equation for the dynamical
 variable $E_2$ contains, besides a noise term, an  effective
 potential that scales like $1/E_2$ and that involves
 the other variable $\xi$.  This effective potential diverges
 at $E_2 = 0$ and constrains the energy to stay positive:
 we  do not need anymore to impose  this physical condition 
 arbitrarily.

\subsection{Analytical results}

\begin{figure}[t]
 \hfill
 \includegraphics*[width=0.4\textwidth]{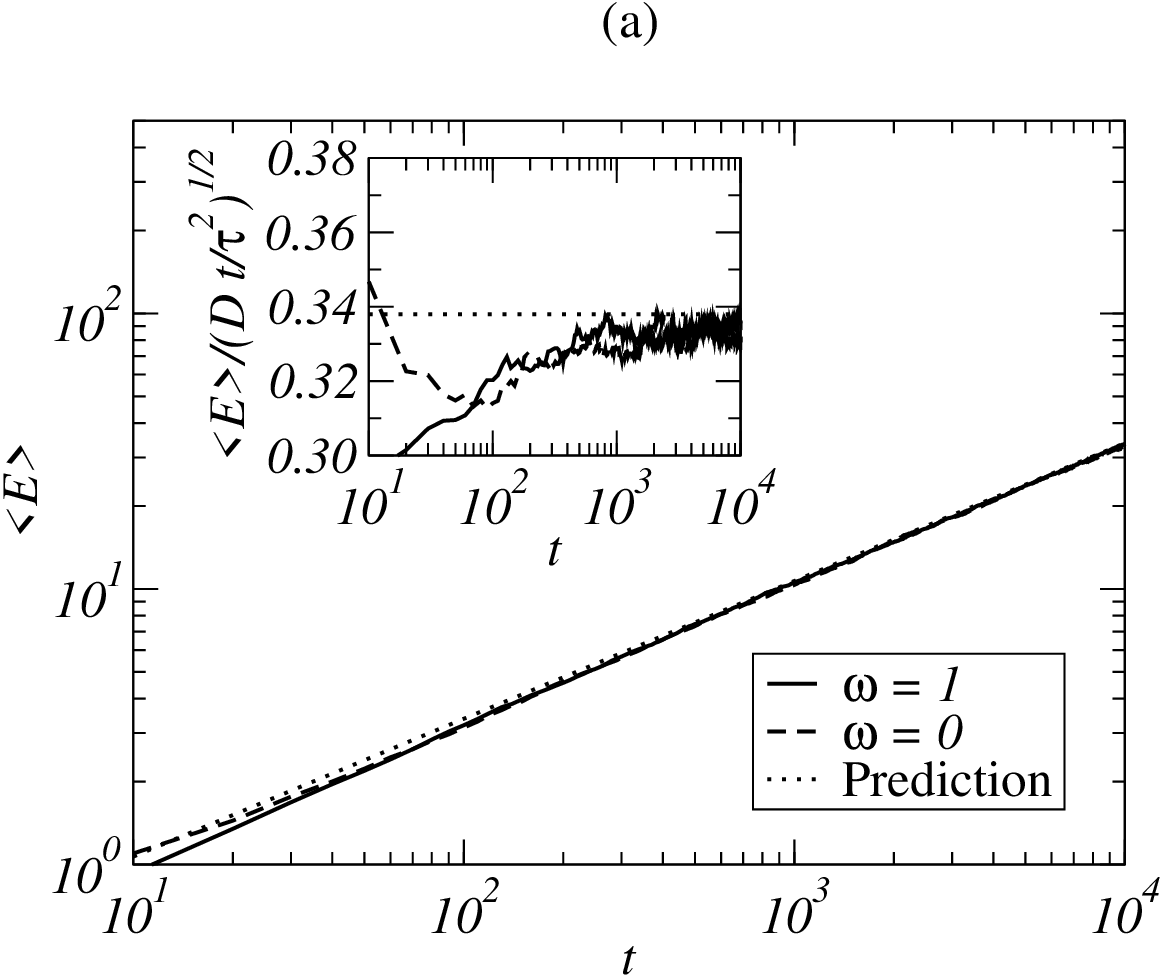}
 \hfill
 \includegraphics*[width=0.38\textwidth]{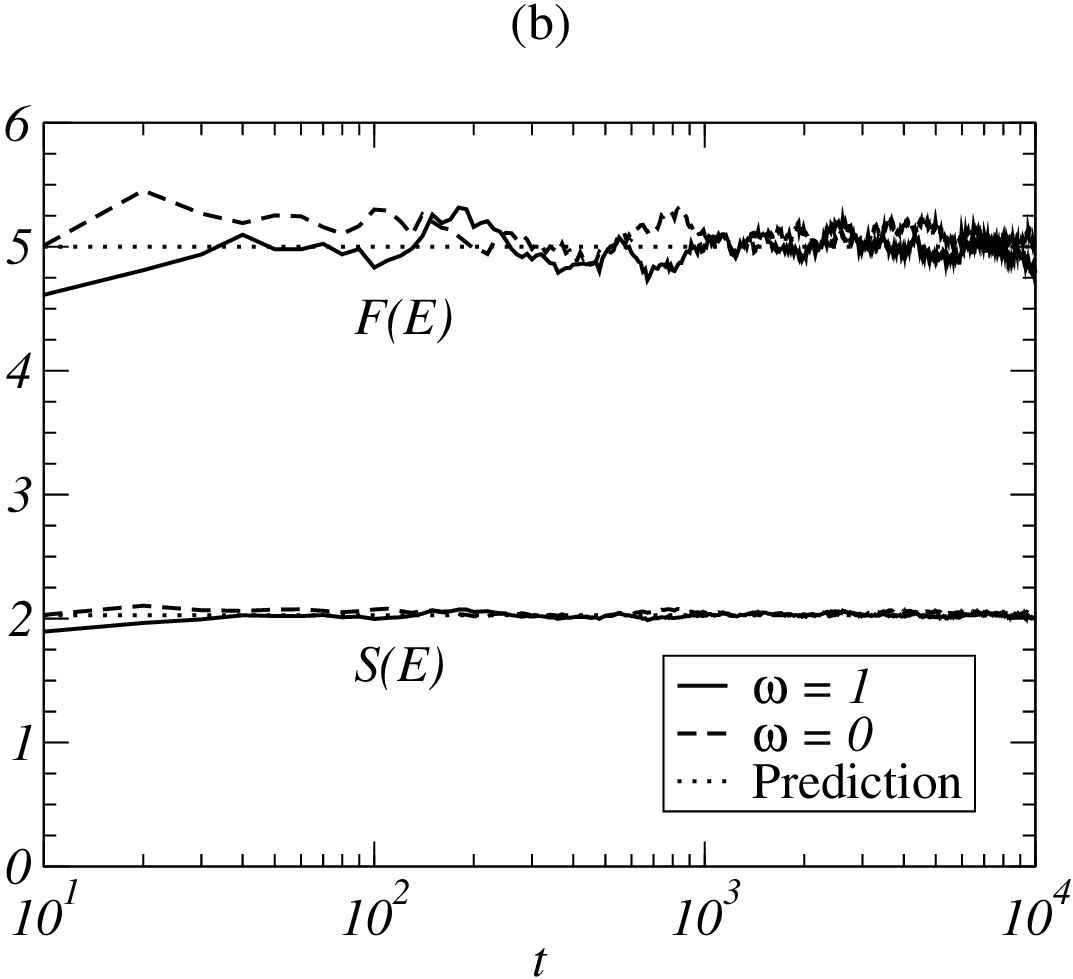}
 \hfill

    \caption{\label{fig:color} Stochastic pendulum with 
Ornstein-Uhlenbeck colored noise: 
Eqs.~(\ref{dyncolOmega}--\ref{OU}) 
are integrated numerically for ${\mathcal D} = 1$, $\tau = 1$. 
Ensemble averages 
are computed over $10^4$ realizations. For numerical values of the pulsation
$\omega = 1.0$ and $0.0$, we plot: $(a)$ the average 
$\langle E \rangle$ and the ratio 
$\langle E \rangle/(\mathcal{D} \,t/\tau^2)^{1/2}$ 
(inset),  $(b)$ the skewness and flatness factors of $E$ \emph{vs.}  time $t$. 
The asymptotic behavior of all observables presented agrees with 
Eqs.~(\ref{eq:scalEcolor}--\ref{eq:flatEcolor}) 
(dotted lines in the figures), irrespective of the value of $\omega$.
}
\end{figure}

 In the coupled system~(\ref{average2},\ref{ZOU}),  the fast angular variable
 has been eliminated and the dimensionality of the original
 problem has been reduced by one. However, Eq.~(\ref{average2})
 is  nonlinear and is 
  not exactly solvable. Nevertheless, 
  the long  time behavior  of $E_2$ can be
 deduced   from the following reasoning.
 Recalling   that $\xi^2$  has a finite mean value,
 equal to ${\mathcal D}/{2\tau}$,  we rewrite the
 evolution equation~(\ref{average2})   of $E_2$  as 
\begin{equation}
 \dot  E_2  =   - \frac{{\mathcal D}} { 8\tau^2 E_2}+  \frac{1}{\sqrt{2}\tau} \eta_2(t)
 -\frac{1} { 4\tau  E_2} \left( \omega^2 \xi +
  \xi^2 -     \langle\xi^2  \rangle   \right)   \,.
 \label{LangeffZ}
\end{equation}
 This Langevin  equation contains two independent noise contributions :
  a white noise  $\eta_2(t)$, 
 and a  (non Gaussian) colored noise,
   $(\omega^2 \xi +  \xi^2 -  \langle\xi^2  \rangle )$, 
 of zero mean value and  of  finite variance.
 This colored noise   is multiplied by a prefactor 
  proportional to $1/E_2$ and,  because  $E_2$ goes  to infinity with time,
  it   becomes negligible  in the  large time  limit in comparison 
 with  the white noise term. Thus,  Eq.~(\ref{LangeffZ}) 
 reduces  asymptotically  to
\begin{equation}
 \dot  E_2  = - \frac{{\mathcal D}} { 8\tau^2 E_2} 
 +  \frac{1}{\sqrt{2}\tau} \eta_2(t) \,.
\label{average2eta}
\end{equation}
 The variables  $E_2$ and $\xi$  are  decoupled at large times:
 the effective problem   is thus   one-dimensional.
In the long time limit,  the variable   $E_2$  is identical
 to the energy  $E$ up to finite terms. We thus  obtain the asymptotic
P.D.F. of the pendulum's energy by explicitely  solving
  the   Fokker-Planck associated  with  Eq.~(\ref{average2eta})
\begin{equation}
  \label{eq:pdfEcolor}
 {\tilde P_t}(E) = 
 \frac{ 2 \sqrt{\tau} }{  \Gamma\left(\frac{1}{4} \right) 
   ({\mathcal D} t)^{1/4} }   E^{ -\frac{1}{2}}  
  \exp\left( -\frac{\tau^2 E^2}{ {\mathcal D} t} \right)   \, .
\end{equation} 
Hence,  $E$ is not a 
  Wiener process on a half line:   its P.D.F. is not  a simple Gaussian.
 From Eq.~(\ref{eq:pdfEcolor}),  we calculate 
 the first two moments of the energy
\begin{eqnarray}
  \label{eq:scalEcolor}
  \langle E \rangle &=& \frac{\sqrt{2} \pi}{\Gamma(\frac{1}{4})^2}
\left(\frac{ {\mathcal D} t}{\tau^2} \right)^{1/2} \simeq
0.338 \left(\frac{ {\mathcal D} t}{\tau^2} \right)^{1/2} \, , \\
  \langle E^2 \rangle &=& \frac{1}{4} \frac{ {\mathcal D} t}{\tau^2} \, . 
  \label{eq:scalE2color}
\end{eqnarray}
Besides  the skewness and  the flatness factors are 
\begin{eqnarray}
  S(E) &=& 
 \frac{\Gamma(\frac{7}{4}) \; \Gamma(\frac{1}{4})}{\Gamma(\frac{5}{4})^{3/2}}
 =  \frac{6\sqrt{2} \pi}{\Gamma(\frac{1}{4})^2}
 \simeq  2.028\ldots,   \label{eq:skewEcolor}\\
  F(E) &=& 
 \frac{\Gamma(\frac{9}{4}) \; \Gamma(\frac{1}{4})}{\Gamma(\frac{5}{4})^2}
= 5.   \label{eq:flatEcolor}
\end{eqnarray}
 The functional dependence of the moments   on  time $t$
  and   on the  parameters   $ {\mathcal D}$ and $\tau$
 is  the same as that obtained in   section~\ref{sec:1staveraging}.
 But the prefactors, which are absolute numbers, are different
 (compare  Eqs.~(\ref{diff1},\ref{quad1}) 
 with  Eqs.~(\ref{eq:scalEcolor},\ref{eq:scalE2color})).
 Numerical simulations of the dynamical
 equations  (\ref{dyncolOmega}, \ref{dyncoltheta} and \ref{OU}), 
  shown in Figs.~\ref{fig:collapse} and \ref{fig:color}, 
    agree  quantitatively   with 
 the  predictions of Eqs.~(\ref{eq:scalEcolor}-\ref{eq:flatEcolor}).
 In particular, we notice  that  the asymptotic   P.D.F. ${\tilde P_t}(E)$,
 and therefore the moments of the energy,
 do  not depend on the value of the mean frequency  $\omega$
 (see  Fig.~\ref{fig:color}).

 Our averaging technique
   thus provides  sound 
 asymptotic results for  the energy of the stochastic pendulum:
  this technique not only   yields  the  correct
 scalings but also leads  to  analytical formulae for
 the large time behavior of the energy.
 This averaging method  could  be carried over  
 to the third  order   to calculate the  subdominant
 corrections to the P.D.F. of the energy. However,  we shall not
 pursue this course any  further:   the calculations become
  very unwieldy and the agreement between
 the analytical  results  and the numerical computations
 is already very satisfactory at  the second order.

\begin{figure}[t]
\centerline{ \includegraphics*[width=0.4\textwidth]{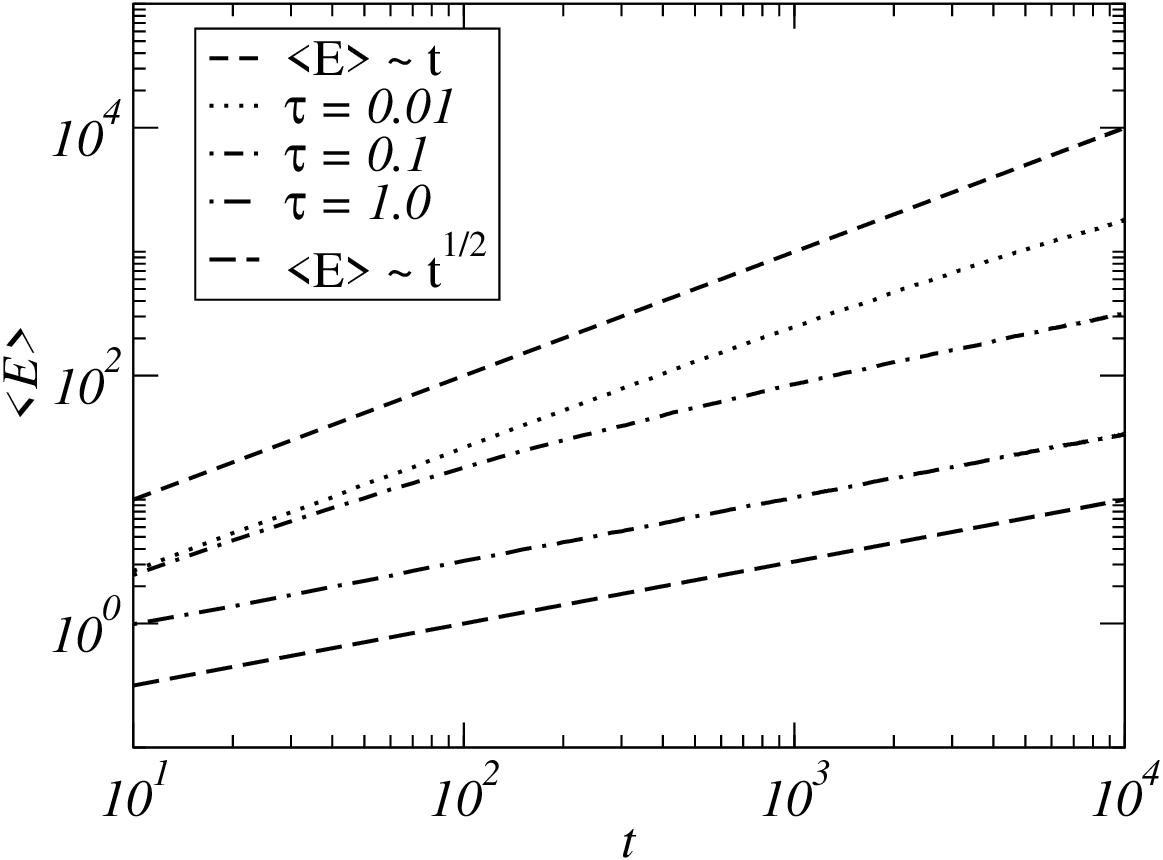}}

    \caption{\label{fig:crossover} Stochastic pendulum with 
Ornstein-Uhlenbeck colored noise: 
Eqs.~(\ref{dyncolOmega}--\ref{OU}) 
are integrated numerically for $\omega = 0.0$, ${\mathcal D} = 1.0$ and
$\tau = 0.01, 0.1$ and $1$. 
Ensemble averages are computed over $10^4$ realizations. 
We plot the average energy  \emph{vs.}  time.
The crossover between white noise behavior ($\langle E \rangle \sim t$)
and colored noise behavior ($\langle E \rangle \sim t^{1/2}$) occurs
later for smaller correlation time $\tau$.
}
\end{figure}

 We emphasize that the   white noise 
 and the colored noise cases  fall in  two distinct  universality
 classes because  the long  time scaling exponents are different. 
 The  colored noise scaling  will always  be observed after 
 a sufficiently  long  time  provided that 
  the correlation time $\tau$ is non-zero.
  However, the effect of this  
 correlation time  appears only when  the period  $T$ of the pendulum 
  is less than  $\tau$.
 This period, which is proportional to $\Omega^{-1}$,
  decreases with time.  At  short times, the period 
  $T$ is  much greater than 
  $\tau$  and  white noise scalings are observed. 
  At large times,  $T \ll  \tau$ and colored noise scalings 
  are satisfied. 
 The crossover time $t_c$ is reached when  $ T \sim \Omega^{-1} \sim \tau$;  
 using   Eq.~(\ref{eq:scalEblanc}), 
 which is  valid for $t < t_c$,  we obtain 
\begin{equation}
 t_c \sim  ( {\mathcal D}\tau^2)^{-1} \, .
\label{tpcross}
\end{equation}
 Hence, when the correlation time $\tau$ becomes vanishingly small,
 the crossover time diverges to infinity and the colored noise
 regime is not reached (simulation times
  needed  to observe the colored noise scalings become increasingly
  long). In Fig.~\ref{fig:crossover}, we plot the behavior of
 the mean energy  {\it vs} time for  ${\mathcal D} = 1.0$ and
$\tau = 0.01, 0.1$ and $1$.  For $\tau = 1$, the colored noise
 scaling regime is obtained  from the very beginning and 
 the curve for $\langle E \rangle$ has a slope 1/2 in
 log-log scale. For $\tau = 0.1$, at short times
 $\langle E \rangle \propto t$ whereas at long times
 $\langle E \rangle \propto t^{1/2}$. The crossover
 is observed around $t_c \sim  100 = \tau^{-2}$.
  For $\tau = 0.01$, the curve for $\langle E \rangle$ has a slope 1
  and the colored noise regime is not reached in this simulation
  $(t_c \sim  10^4).$

 The averaging  method  provides  analytical results
 for  $t  \gg t_c$ and $ t  \ll t_c$. The intermediate
 time regime  is   described by a crossover function of the scaled
 variable $t/t_c$ \cite{philkir2}, which cannot be analyzed  by our technique.
 In the next section, we compare our method 
 with   two  
 well-known approximations  based on effective colored 
 noise Fokker-Planck equations. One of  these  approaches  
  will enable us to derive   approximate formulae  for  $t_c$ and for 
 the  crossover function.

  \section{Comparison with   other  approximation schemes}

     The main difficulty
 for the study of a Langevin equation with colored noise
 stems from its non-Markovian character \cite{vankampen}:
 there exists  no closed  Fokker-Planck equation that  describes
 the evolution of the P.D.F. of the dynamical variables.
 The process can be embedded into a Markov system if the colored  noise
 itself is treated as a variable. However, this  mathematical trick  increases
 the dimension of the problem by one and the noise has to be integrated
 out in the end:   exact calculations can be carried out
 with this method  only in the case of  linear problems. 

  A  colored noise  master equation can be  rigorously 
 derived ({\it e.g.},  with the help of
  functional methods) but it
  involves  correlations between the
   dynamical variables and the noise \cite{hanggirev1,hanggirev2}.
 The equation of motion
 for these  correlations  involves higher order correlations,
 and so on.
 Since this  hierarchy must be stopped  at some stage, 
 this question is
 a genuine closure problem.
  Many different approaches have been 
  devoted to derive   effective Fokker-Planck equations, such as 
 short correlation time expansions 
\cite{lindcol1,lindcol2,sancho1,sancho2},
 unified colored noise approximation \cite{schimansky}, 
 projection methods \cite{fox} and 
  self-consistent decoupling Ans\"atze
  \cite{moss}  (for a general review see \cite{hanggirev3}).

 In section \ref{sec:color}, to  derive  the long time
 behavior of the stochastic pendulum with colored noise, we did not use 
 any closure approximation but  started from 
 the exact Fokker-Planck equation  for the system, 
 treating  the noise as an auxiliary variable.   Thus, 
 we did  not make any   hypothesis on $\tau$ and  our analytical
 formulae  are valid for  any value of the correlation time (for $t$ 
 larger  than the crossover time given in Eq.~(\ref{tpcross})).
   In this section,
 we compare  our results   with  those that  can be 
 derived from some well-known 
 approximation schemes.

  \subsection{Small correlation time   expansion}

 An  approximate  evolution  equation for   the P.D.F. of 
  a Langevin equation with colored noise can be derived 
  in  the case of  short correlation times
 ({\it  i.e.},  in  the white noise limit)  by  expanding 
 the colored noise  master equation around the Markovian point 
\cite{sancho2,hanggirev3}. This procedure leads 
 to a   Fokker-Planck type  equation, with
 effective drift and diffusion coefficients. 
  Applying to our 
 system~(\ref{dyncolOmega}, \ref{dyncoltheta}, \ref{OU})
  the small $\tau$ expansion   derived in \cite{sancho2} 
  for  arbitrary   stochastic equations
  with colored noise, we obtain,    at  first order in   $\tau$, 
the  effective Fokker-Planck equation  for $ P_t(\theta, \Omega)$ 
 \begin{equation}
\frac{ \partial   P_t }{\partial t}  = 
 -\frac{ \partial   }{\partial \theta} \left( \Omega  P_t \right) 
 -   \frac{3 {\mathcal D \tau }}{2}\sin \theta \cos \theta  
  \frac{ \partial  P_t    }{\partial \Omega}
+  \frac{ {\mathcal D}(1 - \tau\Omega)    }{2} 
 \sin^2\theta \frac{ \partial^2   P_t }
 {\partial \Omega^2} 
+ \frac{ {\mathcal D}\tau  }{2}   
\frac { \partial^2 } {\partial\theta  \partial \Omega}
\left( \sin^2\theta  P_t \right) 
   \,.
\label{FPapprox1}
\end{equation}
To simplify the discussion,  we have  taken   the mean frequency $\omega$ 
 equal to  zero.  After integrating  out the rapid
 variations of $\theta$  an averaged Fokker-Planck equation  is obtained 
  which is  similar to Eq.~(\ref{moyblanc}), but with   an effective
 diffusion given by
 ${\mathcal D}_{{\rm eff}} =
 {\mathcal D}(1 - \tau\Omega)$ (because ${\mathcal D}_{{\rm eff}}$
  becomes negative for large 
 values of $\Omega$,  Eq.~(\ref{FPapprox1}) 
  is  valid  only  over  the restricted  region
 of positive diffusion). From dimensional analysis, 
  we  observe that such an  effective
   Fokker-Planck equation  leads  to 
 a normal diffusive behavior, $\Omega \sim t^{1/2}$ and
 $E \sim t$, and therefore  cannot   
 account for the results we have obtained.

 \subsection{Best  Effective Fokker-Planck equation}
   
  We now turn to another small  $\tau$  approximation, which intends to
 improve the first-order effective Fokker-Planck 
   equation~(\ref{FPapprox1}) by  
   summing  contributions
 of the type ${\mathcal D} \tau^n$ (where $n$ is an integer
 larger than 1). The resulting 
 equation has been christened `Best 
 Fokker-Planck  Equation' (B.F.P.E.)  by its proponents
 \cite{lindcol1,lindcol2,lindcol3}. Although this approach 
 is not free from drawbacks  and is known to lead in some cases
 to unphysical results \cite{marchesoni}, 
  we  show  that for the system studied in this work, the
 B.F.P.E. leads to results that agree   with ours.

   An approximate evolution equation for the
 P.D.F. $ P_t(\Omega,\theta)$  is given by the 
  second-order cumulant expansion \cite{vankampen}  of the
 (stochastic)  Liouville equation associated with 
 Eqs.~(\ref{dyncolOmega} and  \ref{dyncoltheta}): 
\begin{equation}
\frac{ \partial   P_t }{\partial t}  = 
 {\bf L}_0 P_t  + \int_0^t dx \langle {\bf L}_1(t) \exp({\bf L}_0x)
    {\bf L}_1(t-x)   \exp(-{\bf L}_0x) \rangle P_t   
\label{FPcumul}
\end{equation}
where  the differential operators are defined as 
 \begin{eqnarray}
    {\bf L}_0  P_t &=&  - \frac{\partial}{\partial \theta}( \Omega P_t )  \, ,
\label{defL0}  \\
    {\bf L}_1(t) P_t &=&  - \frac{ \partial}{\partial \Omega}
 ( \xi(t) \sin\theta \,  P_t) \,.\label{defL1}
  \end{eqnarray}
(Here again we   take $\omega =0$).
   In the Appendix,   we evaluate  the  right hand side of 
 Eq.~(\ref{FPcumul}) and derive,   in the limit 
 $t \to \infty$,   the  following  B.F.P.E. for
 the classical pendulum with colored multiplicative noise
\begin{equation}
 \frac{ \partial   P_t }{\partial t}  =  
 - \frac{\partial}{\partial \theta}( \Omega P_t )
  +  \frac{ {\mathcal D}  }{2 } 
   \frac{\partial^2}{\partial \Omega^2} 
\left(  \frac{ \sin^2\theta - \tau\Omega\sin\theta \cos\theta}
 { 1 + (\tau\Omega)^2} P_t  \right)
  + \frac{ {\mathcal D} \tau }{2 }
 \frac{\partial}{\partial \Omega} \sin\theta
\frac{\partial}{\partial \theta}
  \left( \frac{ ( 1 - (\tau\Omega)^2 ) \sin\theta 
  - 2 \tau\Omega \cos\theta}
   { \left(  1 + (\tau\Omega)^2  \right)^2} 
 P_t  \right) \, . 
   \label{bfpe}
\end{equation}
  Integrating   out   the fast variable $\theta$, we obtain 
 an averaged B.F.P.E  for ${\tilde P}_t(\Omega)$,
the  probability  distribution of 
 the slow variable, 
  \begin{equation}
  \frac{ \partial   {\tilde P}_t }{\partial t}  = 
  \frac{ {\mathcal D}  }{4}
  \frac{\partial}{\partial \Omega} \Big(  \frac{ 1 }
  { 1 + (\tau\Omega)^2}
   \frac{\partial  {\tilde P}_t    }{\partial \Omega}  \Big)  \, .
   \label{avbfpe}
\end{equation}
 For $\tau =0$,  this equation is identical to  the averaged white
 noise Fokker-Planck equation~(\ref{moyblanc}).  For a  non-zero
 correlation time, this equation  predicts  correctly  that $\Omega$
 grows as $t^{1/4}$: this  is  straightforward  from  
 scaling. The crossover between
 white and colored noises  is observed for  $\Omega \sim 1/\tau$
 and this is consistent with the discussion that led to  Eq.~(\ref{tpcross}).
  Although ${\tilde P}_t(\Omega)$ is not  explicitely  calculable, 
   Eq.~(\ref{avbfpe}) implies the following identity 
   \begin{equation}
  \frac{\tau^2}{4} \langle \Omega^4 \rangle  +
    \frac{1}{2} \langle \Omega^2 \rangle =  \frac{ {\mathcal D}  }{4} t \, .
\label{idquadr}
\end{equation}
 When $t$ is small,  the quadratic term dominates over the quartic term,
 and we recover 
  $\langle \Omega^2 \rangle = 2 \langle E \rangle  \simeq  \frac{ {\mathcal D}  }{2} t,$ 
  in agreement with   Eq.(\ref{eq:scalEblanc}). When $t$ is large, 
    the quartic term is dominant and one deduces that
  \begin{equation} 
  \langle E^2 \rangle = 
  \frac{1}{4} \langle \Omega^4 \rangle 
  \simeq \frac{1}{4}   \frac{ {\mathcal D}  }{\tau^2} t \, .
  \label{quadbfpe}
  \end{equation}
   This result is identical to our   Eq.~(\ref{eq:scalE2color}),  
  which was  validated  by   numerical results 
 (see Fig.~\ref{fig:collapse}).
  The identity (\ref{idquadr}) can also be used to 
    derive  an approximate   scaling function    for the  mean energy.
   Let us  define     the  flatness $\phi$ of ${\tilde P}_t(\Omega)$ as
  \begin{equation}
  \langle \Omega^4 \rangle   = \phi  \; \langle \Omega^2 \rangle^2 \, .
  \label{deflatness}
  \end{equation}
   Rigorously speaking  $\phi$ is  a function of time, but  it remains
   a number of order 1.  For simplicity, let us assume
  that    $\phi$   is constant. 
  Substituting Eq.~(\ref{deflatness}) in  Eq.~(\ref{idquadr}),
   and solving  for  $\langle \Omega^2 \rangle$, we obtain 
  \begin{equation}
   \langle E  \rangle  = \frac{1}{2} \langle \Omega^2 \rangle = 
   \frac{ \sqrt{  {\mathcal D}\tau^2\phi \,  t + 1}  -1  }
   { 2  \tau^2\phi   }   \, .
    \end{equation}
This scaling function  explicitely describes 
 the evolution of the energy of the oscillator as a function of time.
 It contains, in particular,  the  linear behavior at short  time  and the 
 $t^{1/2}$ growth at large time. The crossover
 between these two scaling regimes occurs when
 ${\mathcal D}\tau^2 t \sim 1$, {\it i.e.},  precisely
 at  the crossover time  given in Eq.~(\ref{tpcross}).

 The B.F.P.E. approach  has thus allowed us 
 to derive  short and long  time  scalings and to understand qualitatively
 the evolution of the system at intermediate times.  The B.F.P.E. is derived
 from a second-order cumulant expansion 
 in which  higher order terms,
 that are not Fokker-Planck-like,  are discarded.
 This approximation
 has been criticized   \cite{marchesoni}  because the neglected  
 higher-derivative terms  can be  of the same order 
 as  the terms  that have been retained  in the  B.F.P.E.  However, for the
 stochastic pendulum,  these neglected terms do not change qualitatively 
  the large time behavior of the
 dynamical variables. In contrast, the 
   averaging technique  that we have generalized  here 
 is not based on any {\it a priori}
 expansion and    provides   reliable results
 at least for the stochastic system  studied in this work. 
 Of course, the approach  advocated  here has been developed
 for one  particular problem and does not 
  have the generality and versatility of effective
 Fokker-Planck methods. Nevertheless, we strongly believe that 
 the recursive averaging scheme developed  here for 
 the stochastic pendulum can be extended
 to other nonlinear  one-dimensional systems
 subject to  multiplicative
 or additive colored noise  \cite{philkir4}.

\section{Conclusion}

  A nonlinear pendulum subject to parametric noise  undergoes
 a  noise-induced  diffusion  in phase space. 
  The characteristics of this motion  depend
   on the   nature of the randomness: when the noise is white  
  the energy of the pendulum grows linearly with time,
 whereas  it  varies only  as the square root of time 
  when the noise is colored. This change of behavior is
 due to destructive interference 
 between the displacement of the pendulum and the noise term:
 the effect of  the  colored  noise is partially  averaged out  by
  fast angular variations  as soon as  the  
 period of the pendulum becomes smaller than the correlation
 time $\tau$ of the noise. We have carried out an analytical  study
 of this model   by  defining recursively   new
 coordinates in the phase space   and  averaging 
 out the  fast angular  variable. At  zeroth  order,
  the only information obtained is  that the motion is subdiffusive;
   at first order,  this procedure provides
 the correct scalings;  at second order, a  quantitative 
 agreement with numerical simulations is reached. 

  We emphasize that our method  is  different from
 the usual approximations   that involve effective  Fokker-Planck
 equations in which the colored noise does not appear  as an auxiliary
 variable. Our averaging  procedure  
 integrates out the fast dynamical variable and  leads  to 
  an effective stochastic  dynamics
 for the slow variables with   colored  noise. Whereas 
 usual effective  Fokker-Planck    equations  are valid 
 only for  small noise and  for short correlation times, 
  we do not make any  hypothesis on the amplitude 
 of the noise or on the  correlation time $\tau$.
  However,  the  asymptotic subdiffusive regime is reached 
 earlier  for larger values of  $\tau$. 
 Our results   agree    with those  derived  after averaging 
 the `Best Fokker-Planck Equation' (B.F.P.E.) approximation 
 (which is obtained  from  a summation of a 
   truncated cumulant expansion of the Liouville equation).
 This approximation is also  useful to draw a qualitative
    physical picture of the system
  and allows to calculate approximate crossover functions 
  between  short time  (white noise) and  
  long time (colored noise) regimes. 
 
  The recursive averaging scheme that we have  used here for 
 the stochastic pendulum can be extended
 to other systems coupled with  colored noise. In particular,
 we believe that  thanks to   this method  a precise mathematical analysis 
 of  the long  time  behavior of a nonlinear oscillator
  subject to multiplicative or additive colored noise  
  can be carried out \cite{philkir4}.

\acknowledgements

     We  thank  S. Mallick for his helpful  comments on  the
 manuscript  and F. Moulay for useful discussions.

\appendix

  \section{Derivation  of Eq.~(\ref{bfpe})}
\label{sec:proof}

In this appendix, we  derive  the B.F.P.E. 
 (Eq.~(\ref{bfpe}))  following the procedure  of  \cite{lindcol1,lindcol2}.
 We evaluate   the right hand side of Eq.~(\ref{FPcumul}) 
 by applying  the  following  operator formula
 \begin{equation}
 \exp(A) B \exp(-A) =  B + [A,B] + \frac{1}{2!} [A, [A,B]] + 
 \frac{1}{3!} [A, [A, [A,B]]] + \ldots  \,, 
\label{idmat}
\end{equation}
 with  $ A = {\bf L}_0$   and $B = {\bf L}_1(t-x)$. We thus have to  
  calculate  some  commutators of the two operators 
 ${\bf L}_0$ and  ${\bf L}_1(t-x)$ defined in Eqs.~(\ref{defL0}) and 
  (\ref{defL1}).
  By induction, we derive  the following 
 expression  for the $n$-th  commutator 
\begin{equation}
 T_n = 
[{\bf L}_0,[\ldots,[{\bf L}_0, [{\bf L}_0, {\bf L}_1(t-x)]]\ldots]  =  \xi(t-x)
  \left( \frac{ \partial   }{\partial \Omega}  H_1^{(n)}(\Omega,\theta)
   +  \frac{ \partial   }{\partial \theta}  H_2^{(n)}(\Omega,\theta)    \right)
 \, , \label{defTn}
\end{equation}
where the  functions $H_1^{(n)}$   and $H_2^{(n)}$ satisfy the  recursion relations   
\begin{eqnarray}
   H_1^{(n)} &=& -  \Omega  \frac{ \partial H_1^{(n-1)}}{\partial \theta} \, ,  \\
   H_2^{(n)} &=&   H_1^{(n-1)} 
  -  \Omega  \frac{ \partial H_2^{(n-1)}   }{\partial \theta} \, . 
\label{recurrence}
\end{eqnarray}
The first few  terms   can be calculated explicitely  and we obtain 
 \begin{eqnarray}
 H_1^{(0)}   = -\sin\theta   \,\,\,\, &\hbox{ and }& 
\,\,\,\,  H_2^{(0)}   = 0 \, ,\\
 H_1^{(1)}   = \Omega\cos\theta   \,\,\,\, &\hbox{ and }& \,\,\,\, 
  H_2^{(1)}   =  -\sin\theta\, ,\\
H_1^{(2)}   = \Omega^2\sin\theta   \,\,\,\, &\hbox{ and }& \,\,\,\, 
  H_2^{(2)}   =  2\Omega\cos\theta  \, .
\end{eqnarray} 
 The general solution for  the recursion~(\ref{recurrence})
 is readily found:
 \begin{eqnarray}
   H_1^{(n)} &=&  (-1)^{n-1} \Omega^n \sin(\theta + n \frac{\pi}{2})    \, ,  \\
   H_2^{(n)} &=&  n  H_1^{(n-1)} = 
 (-1)^{n-1} n  \Omega^{n-1} \cos(\theta + n \frac{\pi}{2})  
  \, . 
\label{solrecurrence}
\end{eqnarray}
 From Eqs.~(\ref{idmat}) and (\ref{defTn}),   we deduce the following identity
\begin{eqnarray}
\langle {\bf L}_1(t) \exp({\bf L}_0x) 
    {\bf L}_1(t-x)   \exp(-{\bf L}_0x) \rangle 
 =  &\sum_{n=0}^{\infty}&  \frac{x^n}{n!}  \langle {\bf L}_1(t) T_n  \rangle  
 \nonumber \\
  =  -\frac{\partial}{\partial \Omega} \sin\theta
 &\sum_{n=0}^{\infty}&  \frac{x^n}{n!} \langle \xi(t)\xi(t-x)    \rangle  
 \left(  \frac{\partial}{\partial \Omega}  H_1^{(n)}(\Omega,\theta)
   +  \frac{ \partial   }{\partial \theta}
   H_2^{(n)}(\Omega,\theta) \right)     \, . 
\end{eqnarray}
Substituting  in this equation the expressions~(\ref{solrecurrence})
  of $H_1^{(n)}$ and $ H_2^{(n)}$, and  the autocorrelation
 function~(\ref{correlationOU}) of the Ornstein-Uhlenbeck noise, 
 we  find   that  the right hand side of Eq.~(\ref{FPcumul}) 
 is given by 
\begin{eqnarray}
 \int_0^t {\rm d}x \langle {\bf L}_1(t) \exp({\bf L}_0x) 
     {\bf L}_1(t-x)   \exp(-{\bf L}_0x) \rangle  
 =  -\frac{\mathcal D  }{2 \, \tau} 
   \frac{\partial}{\partial \Omega} \sin\theta  \,
 \left(  \frac{\partial}{\partial \Omega} {\mathcal H}_1(\Omega,\theta,t)
     +  \frac{ \partial}{\partial \theta}
    {\mathcal H}_2(\Omega,\theta,t)  \right)  \, ,
 \end{eqnarray}
where we have defined 
\begin{eqnarray} 
{\mathcal H}_1(\Omega,\theta,t)  &=& 
  \sum_{n=0}^{\infty} \frac{ \int_0^t {\rm d}x  x^n {\rm e}^{-x/\tau } }{n!}
  H_1^{(n)} =  
\sum_{n=0}^{\infty} \frac{ \int_0^{{t}/{\tau}}
 {\rm d}x \,  x^n {\rm e}^{-x } }{n!}
  (-1)^{n-1} \tau^{n+1} \Omega^n \sin(\theta + n \frac{\pi}{2}) \, ,
  \label{defgotH1} \\
 {\mathcal H}_2(\Omega,\theta,t)  &=& 
  \sum_{n=0}^{\infty} \frac{ \int_0^t {\rm d}x  x^n {\rm e}^{-x/\tau } }{n!}
  H_2^{(n)} =  
\sum_{n=0}^{\infty} \frac{ \int_0^{{t}/{\tau}}
 {\rm d}x \,  x^n {\rm e}^{-x } }{n!}
  (-1)^{n-1} \tau^{n+1} n  \Omega^{n-1} \cos(\theta + n \frac{\pi}{2}) \, . 
 \label{defgotH2}
\end{eqnarray}
 In the limit $t \to \infty$, the integral  
$\int_0^{{t}/{\tau}} {\rm d}x \,  x^n {\rm e}^{-x }$ converges
 to $ n!$,   and the series defining  
${\mathcal H}_1$ and ${\mathcal H}_2$  can be calculated 
in a closed form. We finally   obtain
\begin{eqnarray} 
{\mathcal H}_1(\Omega,\theta,\infty)  &=& 
  - \tau\frac{ \sin\theta - \tau\Omega \cos\theta}{ 1 + (\tau\Omega)^2}  \, ,\\
  {\mathcal H}_2(\Omega,\theta,\infty)  &=& 
  - \tau^2 \frac{ ( 1 - (\tau\Omega)^2 ) \sin\theta  - 2  \tau\Omega \cos\theta}
   { \left(  1 + (\tau\Omega)^2  \right)^2}   \, .
\end{eqnarray}
This completes  the proof of Eq.~(\ref{bfpe}).

\end{document}